\pdfoutput=1
\documentclass[sigconf,nonacm,pbalance]{acmart}

\usepackage{booktabs}
\usepackage{graphicx}
\usepackage{amsmath}
\usepackage{enumitem}
\setlist[enumerate]{leftmargin=*,labelsep=0.5em}
\usepackage[ruled,vlined,linesnumbered]{algorithm2e}
\usepackage{tikz}
\usetikzlibrary{arrows.meta,positioning,calc,fit,backgrounds}

\makeatletter
\@ifundefined{hyxmp@parse@acmart}{}{\let\hyxmp@parse@acmart\relax}
\makeatother
\setcopyright{none}
\acmDOI{}
\acmISBN{}
\date{}
\title[ABM-SIRTEM]{ABM-SIRTEM: A Hybrid Agent-Based and Epidemiological Model for Pandemic Response}

\author{Sheryl Paul}
\authornote{Equal contribution}
\affiliation{
   \institution{University of Southern California}
   \city{Los Angeles}
   \country{USA}}

\author{Sam Williams}
\authornotemark[1]
\affiliation{
   \institution{University of Southern California}
   \city{Los Angeles}
   \country{USA}}

\author{Preetom Kumar Biswas}
\authornotemark[1]
\affiliation{
   \institution{Arizona State University}
   \city{Phoenix}
   \country{USA}}

\author{Giulia Pedrielli}
\affiliation{
   \institution{Arizona State University}
   \city{Phoenix}
   \country{USA}}

\author{Jyotirmoy V. Deshmukh}
\affiliation{
   \institution{University of Southern California}
   \city{Los Angeles}
   \country{USA}}

\begin{abstract}
The COVID-19 pandemic has had profound impacts on global health, social structures, and economies. It disproportionately affected lower socioeconomic groups and those reliant on interaction-based jobs. Regulatory bodies faced the challenge of designing policies that preserve public health while limiting disruption to economic stability and productivity. Epidemiological models such as SIR and agent-based models (ABMs) have been used to study disease dynamics and the socioeconomic impacts of disease and interventions. Population-level models often simplify individual heterogeneity, while detailed ABMs can become computationally expensive as the numbers of agents and interactions increase. We propose ABM-SIRTEM, a hybrid model that incorporates occupation categories, economic productivity, and welfare at the individual level while dynamically modeling compliance with government interventions. We calibrate the model against historical positive and negative test counts from four U.S. states and examine the resulting compliance dynamics. This framework provides a basis for studying the interaction between disease spread and socioeconomic behavior in pandemic-response planning.
\end{abstract}

\newcommand{\athomerate}{r_{home}}
\newcommand{\atworkrate}{r_{work}}
\newcommand{\healthstate}{HS}
\newcommand{\economicproductivity}{EP}
\newcommand{\healthrisk}{HR}
\newcommand{\socialhappiness}{MW}
\newcommand{\healthstateofi}{HS^i}
\newcommand{\economicproductivityofi}{\economicproductivity^i}
\newcommand{\socialhappinessofi}{\socialhappiness^i}
\newcommand{\healthriskofi}{\healthrisk^i}
\newcommand{\healthstateofj}{\healthstate^j}
\newcommand{\healthstateofjatt}{\healthstateofj_t}
\newcommand{\complianceofi}{C^i}
\newcommand{\occupationofi}{O^i}
\newcommand{\economicproductivityofiatt}{\economicproductivityofi_t}
\newcommand{\socialhappinessofiatt}{\socialhappinessofi_t}
\newcommand{\healthstateofiatt}{\healthstateofi_t}
\newcommand{\healthriskofiatt}{\healthriskofi_t}
\newcommand{\complianceofiatt}{\complianceofi_t}
\newcommand{\economicscore}{E_s}
\newcommand{\welfare}{W}
\newcommand{\welfareofi}{\welfare^i}
\newcommand{\welfareofiatt}{\welfareofi_t}
\newcommand{\meancomplianceatt}{{\bar{C}^g}_t}
\newcommand{\meancomplianceattnext}{{\bar{C}^g}_{t+1}}

\newcommand{\totalmeancompliance}{\bar{C}_t}
\newcommand{\totalmeancompliancenext}{\bar{C}_{t+1}}

\newcommand{\govtrestriction}{R^{RP}_t}
\newcommand{\necessaryinteractionsi}{NI^i_t}
\newcommand{\voluntaryinteractionsi}{VI^i_t}
\RestyleAlgo{ruled}
\SetKwInput{KwInput}{Input}
\SetKwInput{KwOutput}{Output}
\DeclareMathOperator*{\argmax}{arg\,max}
\definecolor{abmblue}{RGB}{31,90,150}
\definecolor{abmfill}{RGB}{232,240,250}
\definecolor{sirgray}{RGB}{90,90,90}
\definecolor{sirfill}{RGB}{238,238,238}
\definecolor{necred}{RGB}{200,40,40}
\definecolor{volgreen}{RGB}{40,130,60}
\definecolor{essc}{RGB}{31,90,150}
\definecolor{remc}{RGB}{60,160,190}
\definecolor{nonc}{RGB}{225,140,40}
\definecolor{updfill}{RGB}{31,90,150}

\tikzset{
  >={Stealth[length=4pt,width=3.5pt]},
  every picture/.style={line width=0.6pt, font=\footnotesize},
  box/.style={draw=abmblue, fill=abmfill, rounded corners=1.5pt, line width=0.6pt,
              inner sep=3pt, align=center},
  gbox/.style={draw=sirgray, fill=sirfill, rounded corners=1.5pt, line width=0.6pt,
               inner sep=3pt, align=center},
  wbox/.style={draw=black!70, fill=white, rounded corners=1.5pt, line width=0.6pt,
               inner sep=3pt, align=center},
  dbox/.style={draw=black!70, dashed, fill=white, rounded corners=1.5pt, line width=0.6pt,
               inner sep=3pt, align=center},
  ubox/.style={draw=updfill, fill=updfill, text=white, rounded corners=2pt, line width=0.6pt,
               inner sep=3pt, align=center, font=\footnotesize\bfseries},
  container/.style={draw=abmblue, fill=abmfill!60, rounded corners=3pt, line width=0.7pt,
                    inner sep=5pt},
  gcontainer/.style={draw=sirgray, fill=sirfill!70, rounded corners=3pt, line width=0.7pt,
                     inner sep=5pt},
  lab/.style={font=\footnotesize, align=center, inner sep=1.5pt},
  ilab/.style={font=\footnotesize\itshape, align=center, inner sep=1.5pt},
  arr/.style={->, line width=0.6pt},
  darr/.style={->, dashed, line width=0.6pt},
  ttl/.style={font=\footnotesize\bfseries, inner sep=2pt},
  person/.pic={
    \fill[pic actions] (0,0.27) circle (0.085);
    \fill[pic actions, rounded corners=2pt] (-0.13,0.15) -- (0.13,0.15) -- (0.13,-0.03) -- (-0.13,-0.03) -- cycle;
  },
}

\hypersetup{pdflang={en-US}}
\begin{document}
\maketitle
\hypersetup{pdfauthor={Sheryl Paul, Sam Williams, Preetom Kumar Biswas, Giulia Pedrielli, Jyotirmoy V. Deshmukh}}
\pagestyle{plain}
\thispagestyle{plain}

\section{Introduction}
The COVID-19 pandemic, caused by severe acute respiratory syndrome
coronavirus 2 (SARS-CoV-2), resulted in widespread mortality, long-term health
impairments, and pressure on healthcare systems
\cite{ourworldindata2024coronavirus,shi2020overview}. It also disrupted economies
and social interactions. These effects were particularly severe for people with
lower incomes and those whose livelihoods depended on in-person interactions
\cite{imf2020covid_response,imf2020great_lockdown}. Governments introduced a wide
range of restrictions and support measures in response to these challenges
\cite{hale2020variation,cheng2020covid}. Understanding the interaction between
disease spread, individual behavior, and economic activity is therefore
important for evaluating pandemic responses.

Models of COVID-19 have been used to study disease spread and the effects of
interventions \cite{covid2021modeling,kucharski2020early}. A common approach uses
population-level compartmental models, such as the SIR (susceptible, infected,
recovered) model, to describe transitions between health categories
\cite{sir1,sir2,sir3}. Extensions of these models can represent additional
disease stages and support the assessment of intervention strategies
\cite{indonesia,exitstrategy}. These models depend on parameter estimation;
simple formulations also assume homogeneous mixing within compartments, which
can limit their ability to represent individual differences.

Agent-based models (ABMs) capture such differences by simulating individual
behaviors and interactions
\cite{abm1,abm2,hunter2022validating,cuevas2020agent,kerr2021covasim,shamil2021agent,silva2020covid,krivorotko2022agent}.
They can represent heterogeneity in susceptibility, social behavior, and contact
patterns, and can be used to assess social distancing and economic effects.
Detailed ABMs can require substantial data and computation, while sensitivity
to parameter choices complicates validation against observations. Methods based
on stochastic gradient descent and automatic differentiation can integrate ABMs
with deep learning frameworks to learn parameters from heterogeneous data
\cite{aamasabm}. Hybrid agent-based and population-level models also provide a
way to combine individual detail with aggregate disease dynamics
\cite{hunter2020hybrid}. Our work focuses on coupling these dynamics with
occupation-dependent economic incentives and the evolution of compliance with
government restrictions.

\noindent\textbf{Contributions.} We propose ABM-SIRTEM, a method that combines
population-level simulation of disease spread with agent-based analysis of
behavior and welfare. Its main features are as follows:
\begin{enumerate}[leftmargin=*,labelsep=0.5em,nosep]
    \item We model occupation types and economic payoffs that depend on
    agents' occupations.
    \item We model the effects of government public health restrictions on
    agents' mobility and contact patterns.
    \item We introduce a quantitative notion of \emph{compliance} to represent
    differences in how individuals respond to regulatory restrictions.
    \item We model the evolution of compliance as agents respond to economic
    incentives, social interactions, and perceived health risks. These dynamics
    provide a mechanism for studying the relationship between restrictions,
    public health, and economic productivity.
    \item We calibrate time-varying transmission rates and compare the resulting
    simulated daily positive and negative test counts with historical COVID-19
    data from four U.S. states.
\end{enumerate}

\begin{figure}[t]
    \centering
    \includegraphics[trim=3cm 0.5cm 1.7cm 2.25cm,clip,width=\linewidth]{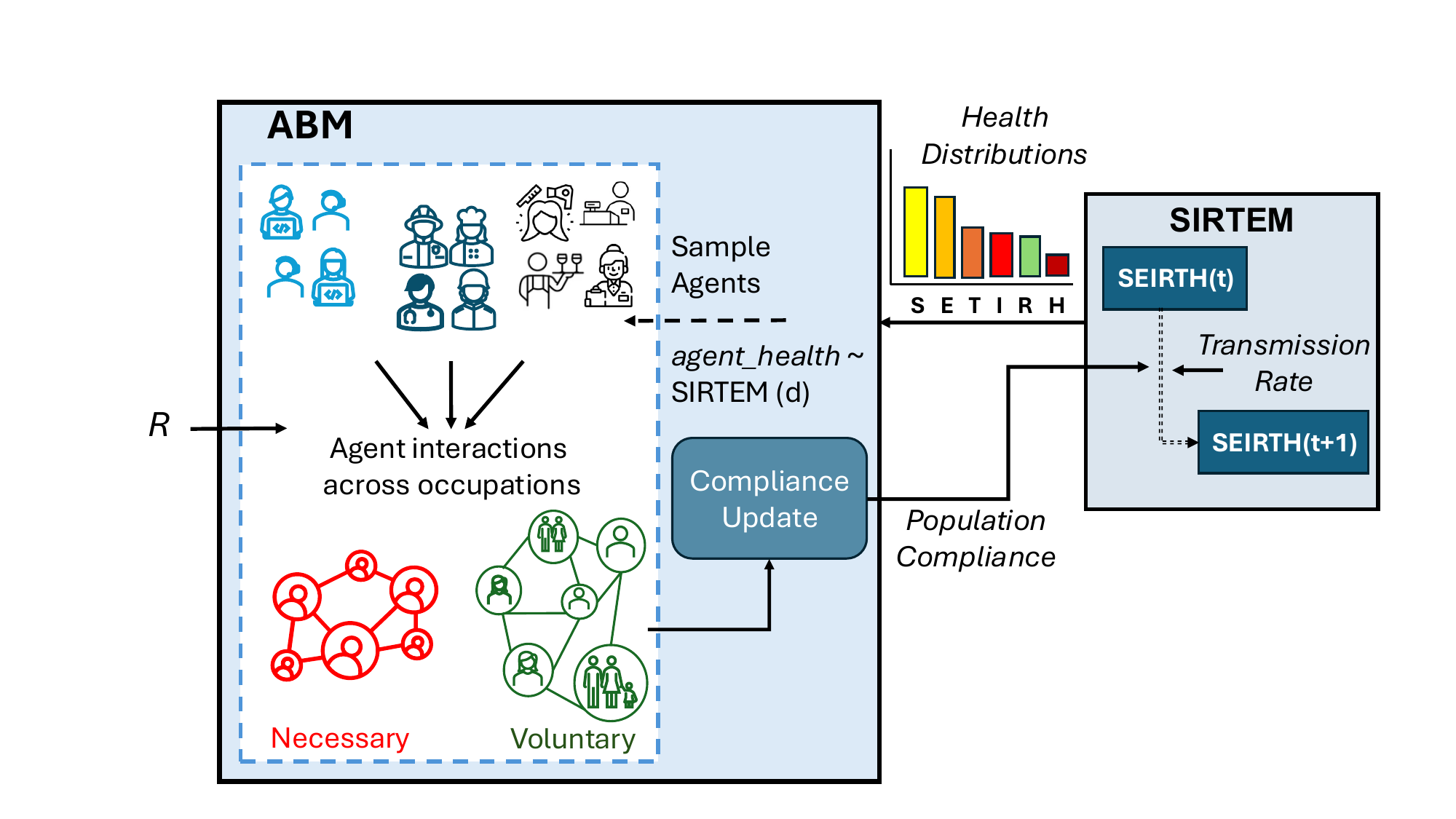}
    \caption{Overview of ABM-SIRTEM. $R$ denotes the government restriction
    index; \texttt{agent\_health} is sampled from the health distribution $d$
    produced by SIRTEM. SEIRTH denotes the health categories in
    Table~\ref{tab:health_states}.}
    \Description{SIRTEM provides a distribution of health states to the ABM.
    The ABM simulates interactions under government restrictions and updates
    compliance, which feeds back into SIRTEM through its infection rate.}
    \label{fig:model-overview}
\end{figure}

\noindent\textbf{Model Overview.}
\begin{enumerate}[leftmargin=*,labelsep=0.5em]
    \item The integrated ABM-SIRTEM model (Fig.~\ref{fig:model-overview}) combines
    SIRTEM \cite{sirtem}, a population-level epidemiological model, with an ABM
    that simulates agent interactions. SIRTEM provides the distribution of health
    categories from which agent health states are sampled. These categories are
    susceptible, exposed, infected, tested, hospitalized, and recovered
    (Table~\ref{tab:health_states}).
    \item The ABM represents agents by their occupation, health status, and
    compliance with government restrictions. Interactions depend on mixing rates
    derived from daily contact frequencies. We define \emph{welfare} as a
    combination of economic productivity, mental well-being, and health risk;
    we collectively refer to these components as welfare parameters.
    \item A regulatory body imposes a \emph{restriction index} that affects
    the number of agent interactions.
    \item Agent compliance evolves over time according to a rule motivated by
    evolutionary game theory. Compliance updates depend on the welfare
    associated with agents' behavior within groups defined by occupation
    and health state.
    \item The ABM supplies mean compliance to SIRTEM. This changes the
    infection rate and the simulated daily positive and negative test counts.
\end{enumerate}

\section{Agent-Based Model}
In this section, we describe the agent-based model (ABM), including agent attributes, the restriction index, and the interaction mechanism. Each agent is initialized with parameters that influence its behavior and interactions.

\begin{table}[t]
    \centering
    \small
    \setlength{\tabcolsep}{4pt}
    \renewcommand{\arraystretch}{1.05}
    \begin{tabular}{@{}p{0.20\columnwidth}p{0.51\columnwidth}p{0.18\columnwidth}@{}}
        \toprule
        \textbf{Variable} & \textbf{Description} & \textbf{Domain} \\
        \midrule
        $E_s$ & Economic productivity unit & $(0,\infty)$ \\
        $M_w$ & Mental well-being unit & $(0,\infty)$ \\
        $\gamma$ & Essential-worker risk discount factor & $[0,1]$ \\
        $\alpha$ & Compliance learning rate & $[0,1]$ \\
        $w_{EP}$, $w_{MW}$, $w_{HR}$ & Weights for economic productivity, mental well-being, and health risk & $[0,1]$ \\
        $\sigma^2_{NI}$, $\sigma^2_{VI}$ & Variances of necessary and voluntary interaction counts & $(0,\infty)$ \\
        $\sigma_C^2$ & Variance used to sample compliance & $(0,\infty)$ \\
        $m$ & Population fraction used to scale voluntary interactions & $[0,1]$ \\
        \bottomrule
    \end{tabular}
    \caption{Hyperparameters of the agent-based model}
    \label{tab:abm_hyperparameters}
\end{table}

\begin{table}[b]
    \centering
    \small
    \setlength{\tabcolsep}{4pt}
    \renewcommand{\arraystretch}{1.05}
    \begin{tabular}{@{}p{0.45\columnwidth}p{0.49\columnwidth}@{}}
        \toprule
        \textbf{Variable} & \textbf{Domain or possible values} \\
        \midrule
        Agent set $(N)$ & $\{1,\ldots,|N|\}$ \\
        Occupation $(O)$ & Essential, Remote, Non-essential non-remote \\
        Health state $(HS)$ & $\{S,E,T,I,R,H\}$ \\
        Compliance $(C)$ & $[0,1]$ \\
        Economic productivity $(EP)$ & $[0,\infty)$ \\
        Mental well-being $(MW)$ & $[0,\infty)$ \\
        Health risk $(HR)$ & $(-\infty,0]$ \\
        Welfare $(W)$ & $\mathbb{R}$ \\
        Necessary interactions $(NI)$ & $\necessaryinteractionsi \subseteq N\setminus\{i\}$ \\
        Voluntary interactions $(VI)$ & $\voluntaryinteractionsi \subseteq N\setminus\{i\}$ \\
        Government restriction policy $(\govtrestriction)$ & $[0,1]$ \\
        \bottomrule
    \end{tabular}
    \caption{Agent attributes and parameters of the agent-based model}
    \label{tab:abm}
\end{table}

\subsection{Agent Description}
Let $N$ denote the set of agents, indexed by $i\in\{1,\ldots,|N|\}$. Agent attributes are summarized in Table~\ref{tab:abm}. Each agent belongs to one of three occupation groups: \textit{Essential}, \textit{Remote}, or \textit{Non-essential non-remote}. We denote the set of occupation groups by $O$ and the occupation of agent $i$ by $\occupationofi\in O$. These categories encode the following assumptions about work and compliance:
\begin{itemize}[left=0pt]
    \item \textit{Essential workers} work regardless of government restrictions. Their non-compliance models additional social interactions beyond work-related interactions.
    \item \textit{Remote workers} also work regardless of government restrictions, and their employment does not incur additional interaction-related health risk. Their non-compliance concerns voluntary social interactions.
    \item \textit{Non-essential non-remote workers} benefit economically from non-compliance because additional interactions can represent opportunities to work as well as socialize.
\end{itemize}

\noindent The ABM simulation runs for $T$ daily time periods. In addition to occupation, each agent has the following time-varying attributes:
\begin{enumerate}[leftmargin=*,labelsep=0.5em]
\item \textbf{Health State:} The health state of agent $i$ at time $t$ is denoted by $\healthstateofiatt\in HS$, where $HS=\{S,E,T,I,R,H\}$. These categories correspond to susceptible, exposed, tested, infected, recovered, and hospitalized, respectively.

\item \textbf{Economic Productivity:} The economic productivity $\economicproductivityofiatt\in[0,\infty)$ represents the monetary benefit that agent $i$ receives from employment and interactions at time $t$, such as wages, commissions, and tips.

\item \textbf{Mental Well-being:} The mental well-being $\socialhappinessofiatt\in[0,\infty)$ represents the non-monetary benefit that agent $i$ receives from interactions with other agents at time $t$. Previous studies have examined and attempted to quantify the mental health and social impacts of isolation during lockdowns~\cite{eddy2021social,alradhawi2020covid_mental_wellbeing,biester2020quantifying}. We similarly represent the benefits of social interactions using a mental well-being score proportional to the number of voluntary interactions.

\item \textbf{Health Risk:} The health risk $\healthriskofiatt\in(-\infty,0]$ represents the daily health risk penalty incurred by agent $i$ through interactions at time $t$. In an interaction between agents $i$ and $j$, each agent incurs a penalty based on the other agent's health state. The penalty incurred by agent $i$ when interacting with agent $j$ is
\begin{equation}
\operatorname{Risk}^{ij}_t =
\begin{cases}
-p_S & \text{if } \healthstateofjatt = S,\\
-p_E & \text{if } \healthstateofjatt = E,\\
-p_T & \text{if } \healthstateofjatt = T,\\
-p_I & \text{if } \healthstateofjatt = I,\\
-p_R & \text{if } \healthstateofjatt = R,\\
-p_H & \text{if } \healthstateofjatt = H,
\end{cases}
\label{eq:penalty}
\end{equation}
where $p_R<p_S<p_E\leq p_T\leq p_I<p_H$ are positive constants representing the relative infection risk of interacting with agents in these categories.

\item \textbf{Compliance:} The compliance $\complianceofiatt\in[0,1]$ represents the degree to which agent $i$ conforms to government restrictions by reducing voluntary interactions, defined in Section~\ref{sec:interactions}. For a fixed restriction index, a compliance score of $1$ sets the mean voluntary interaction count to zero, whereas a score of $0$ maximizes that mean.

\item \textbf{Welfare Index:} The welfare $\welfareofiatt\in\mathbb{R}$ of agent $i$ at time $t$ is a weighted sum of economic productivity, mental well-being, and health risk. We use this index as a modeling proxy motivated by research and debate on measuring welfare and subjective well-being~\cite{dolan2012measuring,johns2007happiness,fabian2019racing}. We calculate welfare as
\begin{equation}
\begin{aligned}
\welfareofiatt ={}& w_{EP}\cdot\economicproductivityofiatt
                  + w_{MW}\cdot\socialhappinessofiatt\\
                 &+ w_{HR}\cdot\healthriskofiatt,
\end{aligned}
\label{eq:agent_welfare}
\end{equation}
where $w_{EP}$, $w_{MW}$, and $w_{HR}$ are the respective component weights.\footnote{We use equal weights for the three components.}
\end{enumerate}

\textbf{Restriction Policy.} We represent government restrictions in each time period by a restriction index $\govtrestriction\in[0,1]$, which influences agent interactions. A value of $0$ represents no restrictions, and a value of $1$ represents a full lockdown.

\textbf{Compliance and Health Dynamics.} We use the population-level SIRTEM model, described in Section~\ref{sec:integration}, to simulate disease dynamics, while the ABM simulates population compliance. Let $HS_t$ denote the categorical distribution of health states in the population at time $t$, and let $\totalmeancompliance=\frac{1}{|N|}\sum_{i\in N}\complianceofiatt$ denote mean population compliance. The coupled updates are summarized by
\begin{align}
HS_{t+1} &= \texttt{SIRTEM}(HS_t,\totalmeancompliance),\label{eq:diff_sirtem}\\
\totalmeancompliancenext &= \texttt{ABM}(HS_{t+1},\totalmeancompliance,\govtrestriction),\label{eq:diff_abm}
\end{align}
where \texttt{SIRTEM} runs for one day and \texttt{ABM} performs the iteration detailed in Algorithm~\ref{alg:abm} and Section~\ref{sec:abm_iteration}. We first describe how simulated interactions affect welfare and, in turn, individual compliance.

Let $G=O\times HS$ denote the set of occupation--health groups. For each group $g\in G$, let
\[
N_g=\{i\in N:(\occupationofi,\healthstateofiatt)=g\}.
\]
For the nonempty groups $G_+=\{g\in G:|N_g|>0\}$, mean compliance is
\[
\meancomplianceatt=\frac{1}{|N_g|}\sum_{i\in N_g}C_t^i,
\qquad g\in G_+.
\]
We model the evolution of compliance using a version of the \emph{replicator equation}~\cite{egt_games,egt_stanford}, in which agents mimic the compliance of agents with high welfare. Agents imitate only members of their own occupation--health group because behavior that benefits an agent in another group may lower their own welfare. For example, a non-essential non-remote worker may fare worse by imitating a remote worker whose economic productivity is unaffected by compliance.

For agent $i\in N_g$, the compliance update is
\begin{equation}
\delta C_t^i=\alpha\sum_{j\in N_g}(W_t^i-W_t^j)(C_t^i-C_t^j),
\label{eq:agent_compliance_update}
\end{equation}
where $\alpha$ is a learning rate controlling adaptation. Averaging over agents in a nonempty group gives
\begin{align}
\delta\meancomplianceatt &= \frac{1}{|N_g|}\sum_{i\in N_g}\delta C_t^i,\\
\meancomplianceattnext &= \meancomplianceatt+\delta\meancomplianceatt.
\label{eq:group_compliance_update}
\end{align}

\textbf{ABM State.} Mean population compliance provides the following summary of the ABM state:
\begin{equation}
s_t=\totalmeancompliance=\frac{1}{|N|}\sum_{g\in G_+}|N_g|\,\meancomplianceatt.
\label{eq:population_mean_compliance}
\end{equation}
For a fixed sampled population and its group memberships, the corresponding update is
\begin{equation}
\delta s_t=\frac{1}{|N|}\sum_{g\in G_+}|N_g|\,\delta\meancomplianceatt,
\label{eq:population_mean_compliance_update}
\end{equation}
with $s_{t+1}=s_t+\delta s_t$.

\subsection{Agent Interactions}\label{sec:interactions}
Simulated interactions determine economic productivity, mental well-being, and health risk for each agent $i\in N$. We distinguish necessary interactions $\necessaryinteractionsi$ from voluntary interactions $\voluntaryinteractionsi$. Their numbers depend on the agent's occupation, health state, and compliance, as well as the restriction index at time $t$.

\subsubsection{Classification of Interactions}
\begin{enumerate}[leftmargin=*,labelsep=0.5em]
\item \textbf{Necessary Interactions:} Necessary interactions are unavoidable in daily life, regardless of compliance or government restrictions. They arise from an agent's occupation or home environment. Essential workers must interact at work as well as at home, so their necessary interaction count depends on both the at-home mixing rate $\athomerate$ and the at-work mixing rate $\atworkrate$. For remote and non-essential non-remote workers, only $\athomerate$ is relevant. Let $\necessaryinteractionsi$ denote the set of agents with whom agent $i$ has necessary interactions at time $t$. Its size is sampled according to
\[
|\necessaryinteractionsi|\sim\mathcal{N}(\mu_{NI}^i,\sigma_{NI}^2),
\]
where
\[
\mu_{NI}^i=
\begin{cases}
(\athomerate+\atworkrate)|N| & \text{if }\occupationofi=\text{Essential},\\
\athomerate|N| & \text{otherwise}.
\end{cases}
\]
Thus, the occupation determines the mean interaction count, and $\sigma_{NI}^2$ is the variance.

\item \textbf{Voluntary Interactions:} Voluntary interactions are optional and depend on an agent's compliance and the government restriction index. Let $\voluntaryinteractionsi$ denote the set of agents with whom agent $i$ voluntarily interacts. The parameter $m\in[0,1]$ scales the interaction count relative to the population size. The number of voluntary interactions is sampled according to
\[
|\voluntaryinteractionsi|\sim\mathcal{N}(\mu_{VI}^i,\sigma_{VI}^2),
\]
where
\[
\mu_{VI}^i=m|N|(1-\complianceofiatt)(1-\govtrestriction),
\]
and $\sigma_{VI}^2$ is the variance.
\end{enumerate}

\subsubsection{Interaction Simulation}
Interactions affect welfare according to their type and number and the occupations and health states of the interacting agents.
\begin{enumerate}[leftmargin=*,labelsep=0.5em]
\item \textbf{Economic Productivity:} Economic productivity depends on an agent's occupation and interactions.
\begin{itemize}[leftmargin=*]
\item \textit{Essential and remote workers} can work regardless of government restrictions and receive a fixed payoff determined by the mixing rates, the interaction-scale parameter $m$, and the population size $|N|$.
\item \textit{Non-essential non-remote workers} receive a payoff proportional to their number of voluntary interactions with non-hospitalized agents. The indicator $\vmathbb{1}_{\healthstateofjatt\neq H}$ equals $1$ when agent $j$ is not hospitalized and $0$ otherwise.
\end{itemize}
We define economic productivity as
\begin{equation}
\economicproductivityofiatt=
\begin{cases}
\economicscore(\athomerate+\atworkrate)m|N| &
\begin{gathered}\text{if }\occupationofi\in\\[-2pt]\{\text{Essential, Remote}\},\end{gathered}\\
\economicscore\displaystyle\sum_{j\in\voluntaryinteractionsi}\vmathbb{1}_{\healthstateofjatt\neq H} & \text{otherwise},
\end{cases}
\label{eq:economic_productivity_score}
\end{equation}
where $\economicscore>0$ is the unit of economic gain. Essential and remote workers receive a fixed payoff, while non-essential non-remote workers can increase their payoff through additional voluntary interactions with non-hospitalized agents.

\item \textbf{Mental Well-being:} Each voluntary interaction gives each agent a mental well-being payoff of $M_w>0$. The total is
\[
\socialhappinessofiatt=M_w|\voluntaryinteractionsi|.
\]

\item \textbf{Health Risk Penalty:} A discount factor $\gamma\in[0,1]$ reduces the health risk penalty for essential workers' necessary interactions, reflecting the assumption that these interactions take place with appropriate precautions and protective measures.\footnote{Examples include masking and sanitization requirements in supermarkets and personal protective equipment for healthcare workers.} Using the risk function in Equation~\ref{eq:penalty}, we define
\begin{equation}
\begin{aligned}
\healthriskofiatt={}&\sum_{j\in\voluntaryinteractionsi}\operatorname{Risk}^{ij}_t\\
&+\begin{cases}
\gamma\displaystyle\sum_{j\in\necessaryinteractionsi}\operatorname{Risk}^{ij}_t
& \text{if }\occupationofi=\text{Essential},\\
\displaystyle\sum_{j\in\necessaryinteractionsi}\operatorname{Risk}^{ij}_t
& \text{otherwise}.
\end{cases}
\end{aligned}
\label{eq:health_risk_equations}
\end{equation}
\end{enumerate}

\subsection{ABM Iteration}\label{sec:abm_iteration}
Algorithm~\ref{alg:abm} summarizes an ABM iteration. At time $t$, the ABM receives the health state distribution $HS_{t+1}$ from SIRTEM. In each of $R_b$ rounds, it samples agents stratified into occupation--health groups using $HS_{t+1}$ and the occupation distribution. Compliance scores are sampled from the corresponding group distribution, $\complianceofiatt\sim\mathcal{N}(\meancomplianceatt,\sigma_C^2)$, and bounded to $[0,1]$.

The ABM uses the mixing rates $\athomerate$ and $\atworkrate$ and the restriction index $\govtrestriction$ to sample necessary and voluntary interactions. It then computes each agent's economic productivity, mental well-being, health risk, and welfare, followed by the compliance update in Equation~\ref{eq:agent_compliance_update}. These updates yield the group mean updates in Equation~\ref{eq:group_compliance_update} and the population mean update in Equation~\ref{eq:population_mean_compliance_update}. The updates are averaged over the $R_b$ rounds to obtain updated group means and the updated population mean $\totalmeancompliancenext$, which is fed back into SIRTEM.

Averaging over rounds reduces the variance arising from agent sampling and interaction simulation. The round budget $R_b$ therefore controls the tradeoff between variance and computational cost.

\SetKwFunction{SampleAgents}{SampleAgents}
\SetKwFunction{Mandatory}{SimMandatoryInteractions}
\SetKwFunction{Voluntary}{SimVoluntaryInteractions}
\SetKwFunction{Compliance}{ComputeComplianceUpdate}
\SetKwFunction{MeanCompliance}{PopulationMeanCompliance}

\begin{algorithm}[tb]
\caption{ABM Iteration Overview}\label{alg:abm}
\KwInput{ABM state summary $s_t=\bar{C}_t$\newline Health state distribution $HS_{t+1}$\newline At-home mixing rate $\athomerate$\newline At-work mixing rate $\atworkrate$\newline Government restriction index $\govtrestriction$\newline Round budget $R_b$}
\For{$n=1,\ldots,R_b$}{
    $N\gets\SampleAgents(s_t,HS_{t+1})$\;
    $\Mandatory(N,\govtrestriction,\atworkrate)$\;
    $\Voluntary(N,\govtrestriction,\atworkrate,\athomerate)$\;
    $\delta s_t^n\gets\Compliance(N)$\;
}
$\delta s_t\gets\frac{1}{R_b}\sum_{n=1}^{R_b}\delta s_t^n$\;
$s_{t+1}\gets s_t+\delta s_t$\;
$\totalmeancompliancenext\gets\MeanCompliance(s_{t+1})$\;
\KwOutput{ABM state summary $s_{t+1}$\newline Government restriction index $\govtrestriction$\newline Mean compliance $\bar{C}_{t+1}$}
\end{algorithm}
\section{Integration with SIRTEM}\label{sec:integration}
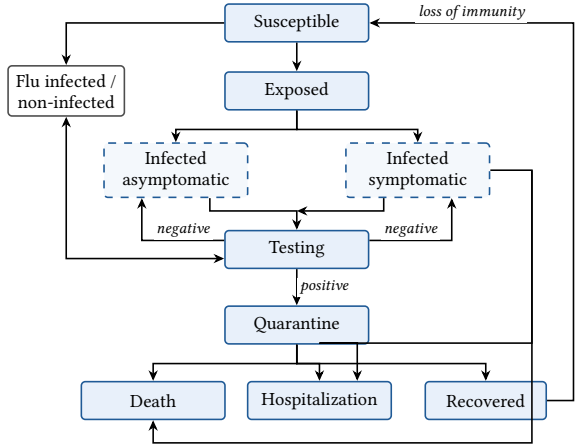
\begin{figure}[t]
    \centering
\begin{tikzpicture}[x=1cm,y=1cm,
  comp/.style={box, minimum height=0.5cm, minimum width=1.9cm, inner xsep=4pt},
  inf/.style={comp, dashed, fill=abmfill!50!white},
  elab/.style={font=\scriptsize\itshape, fill=white, inner sep=1pt}]
  \node[comp] (S)  at (0,0)        {Susceptible};
  \node[comp] (E)  at (0,-0.9)     {Exposed};
  \node[inf]  (IA) at (-1.6,-1.95) {Infected\\asymptomatic};
  \node[inf]  (IS) at (1.6,-1.95)  {Infected\\symptomatic};
  \node[comp] (T)  at (0,-3.0)     {Testing};
  \node[comp] (Q)  at (0,-4.0)     {Quarantine};
  \node[comp] (D)  at (-1.9,-5.0)  {Death};
  \node[comp] (H)  at (0.3,-5.0)   {Hospitalization};
  \node[comp, minimum width=1.6cm] (Rc) at (2.5,-5.0) {Recovered};
  \node[wbox, align=center, minimum height=0.5cm] (F) at (-3.05,-0.9) {Flu infected /\\non-infected};

  \draw[arr] (S) -- (E);
  \draw[arr] (E.south) -- ++(0,-0.25) -| (IA.north);
  \draw[arr] (E.south) -- ++(0,-0.25) -| (IS.north);
  \draw[arr] ($(IA.south)+(0.45,0)$) |- ($(T.north)+(0,0.25)$) -- (T.north);
  \draw[arr] ($(IS.south)+(-0.45,0)$) |- ($(T.north)+(0,0.25)$);
  \draw[arr] ($(T.west)+(0,0.12)$) -| node[elab, pos=0.25, above] {negative} ($(IA.south)+(-0.45,0)$);
  \draw[arr] ($(T.east)+(0,0.12)$) -| node[elab, pos=0.25, above] {negative} ($(IS.south)+(0.45,0)$);
  \draw[arr] (T) -- node[elab, right] {positive} (Q);
  \draw[arr] (S.west) -| (F.north);
  \draw[<->] (F.south) |- ($(T.west)+(0,-0.12)$);
  \draw[arr] (Q.south) -- ++(0,-0.25) -| (D.north);
  \draw[arr] (Q.south) -- ++(0,-0.25) -| (H.north);
  \draw[arr] (Q.south) -- ++(0,-0.25) -| (Rc.north);
  \draw[arr] (IS.east) -- ++(0.55,0) |- ($(H.north)+(0,0.5)$) -- ($(H.north)+(0.5,0.5)$) -- ($(H.north)+(0.5,0)$);
  \draw[arr] ($(IS.east)+(0.55,0)$) |- ($(D.south)+(0,-0.3)$) -- (D.south);
  \draw[arr] (Rc.east) -- ++(0.35,0) |- node[elab, pos=0.75, above] {loss of immunity} (S.east);
\end{tikzpicture}
    \caption{High-level overview of the SIRTEM model. Dashed boxes are the infected compartments; arrows show the transitions between compartments.}
    \Description{SIRTEM health states and transitions through infection, testing, quarantine, hospitalization, and recovery.}
    \label{fig:sirtem-model}
\end{figure}

\begin{table}[b]
    \centering
    \small
    \setlength{\tabcolsep}{4pt}
    \renewcommand{\arraystretch}{1.05}
    \begin{tabular}{@{}ll@{}}
        \toprule
        \textbf{Health state} & \textbf{SIRTEM subcompartments} \\
        \midrule
        Susceptible ($S$) & S, FPS, FPI, FS, GS \\
        Exposed ($E$) & E \\
        Infected ($I$) & PA, PS, IA, IS \\
        Tested ($T$) & AT, ST, QAP, QSP \\
        Hospitalized ($H$) & HBQ, HBT, HDQ, HDT \\
        Recovered ($R$) & KR, UR, IM \\
        \bottomrule
    \end{tabular}
    \par\smallskip
    \begin{minipage}{\linewidth}
        \footnotesize\raggedright
        S: susceptible; FPS: falsely presumed susceptible; FPI: falsely presumed immune; FS: flu symptomatic; GS: general symptoms; E: exposed; PA: pre-asymptomatic; PS: presymptomatic; IA: infected asymptomatic; IS: infected symptomatic; AT: asymptomatic tested; ST: symptomatic tested; QAP: quarantined after an asymptomatic test; QSP: quarantined after a symptomatic test; HBQ: hospitalized before quarantine; HBT: hospitalized before testing; HDQ: hospitalized during quarantine; HDT: hospitalized during testing; KR: known recovered; UR: unknown recovered; IM: immune.
    \end{minipage}
    \caption{Aggregation of SIRTEM subcompartments into ABM health states.}
    \label{tab:health_states}
\end{table}

SIRTEM (Spatially Informed Rapid Testing for Epidemic Modeling)~\cite{sirtem} extends compartmental epidemic models to represent testing, quarantine, hospitalization, and immunity. It distinguishes asymptomatic and symptomatic infections, non-COVID illness, and incorrectly presumed susceptibility or immunity. It also accounts for false-positive and false-negative test results and for spatial differences in population concentrations and mixing rates. Its subcompartments provide a detailed representation of the population, with transitions governed by delay differential equations (DDEs) and their associated rates (Figure~\ref{fig:sirtem-model}).

The exposure process is modeled by
\begin{equation}
\frac{dE}{dt} = \left(\beta S(t) + \beta' FPI(t)\right)\frac{Inf(t)}{N_{\mathrm{pop}}}
- \left(per_a + per_s\right)E(t),
\label{eq:sirtem_exposure}
\end{equation}
where $E(t)$ is the number of exposed individuals at time $t$, $N_{\mathrm{pop}}$ is the epidemiological population size, and $Inf(t)$ is the infectious-population term defined by SIRTEM~\cite{sirtem}. In SIRTEM, this term accounts for differences in infectiousness between symptomatic and asymptomatic individuals. The populations $S(t)$ and $FPI(t)$ are susceptible and falsely presumed immune individuals, respectively, with corresponding infection rates $\beta$ and $\beta'$. Falsely presumed immunity can arise from a false-positive antibody test. The parameters $per_a$ and $per_s$ denote the fractions of infected individuals who are asymptomatic and symptomatic, respectively.

\begin{table}[t]
    \centering
    \small
    \setlength{\tabcolsep}{4pt}
    \renewcommand{\arraystretch}{1.05}
    \begin{tabular}{@{}llc@{}}
        \toprule
        \textbf{Variable} & \textbf{Description} & \textbf{Domain} \\
        \midrule
        $\athomerate$ & At-home mixing rate & $[0,1]$ \\
        $\atworkrate$ & At-work mixing rate & $[0,1]$ \\
        $\beta$ & Infection rate & $[0,1]$ \\
        $per_a$ & Asymptomatic fraction & $[0,1]$ \\
        $per_s$ & Symptomatic fraction & $[0,1]$ \\
        $\phi_s$ & Testing rate & $[0,1]$ \\
        $\lambda_q$ & Quarantine length (days) & $\mathbb{Z}_{>0}$ \\
        \bottomrule
    \end{tabular}
    \caption{Selected SIRTEM model parameters and their stated domains.}
    \label{tab:sirtem_hyperparameters}
\end{table}

The remaining dynamic processes and equations are described in the SIRTEM paper~\cite{sirtem}. We aggregate its subcompartments into the ABM health states listed in Table~\ref{tab:health_states}. SIRTEM also produces daily positive and negative test counts, which we compare with observations when calibrating the integrated model.

The key SIRTEM parameters for our work are the infection rate $\beta$, testing rate $\phi_s$, and quarantine duration $\lambda_q$ (Table~\ref{tab:sirtem_hyperparameters}). The default quarantine duration is 14 days. The original SIRTEM calibration models weekly infection and testing rates using second-order autoregressive, or AR(2), models~\cite{sirtem}. To couple SIRTEM with the mean population compliance $\totalmeancompliance$ produced by the ABM, we introduce a weekly transmission parameter $\tau^{tr}_w$ and compute the daily infection rate as
\begin{equation}
\beta_t = \tau^{tr}_w \cdot \totalmeancompliance.
\label{eq:sirtem_infection_update}
\end{equation}
Here, $\tau^{tr}_w \in \mathbb{R}_{>0}$ is the transmission parameter for week $w$. Let $W=\lceil T/7\rceil$ denote the number of weeks in a $T$-day simulation, and let $\tau^{tr}=(\tau^{tr}_w)_{w=0}^{W-1}$ denote the sequence of weekly transmission parameters. These learned parameters capture virus transmission together with effects of mixing and government restrictions. Because $\totalmeancompliance$ changes daily, we evaluate $\beta_t$ daily using the transmission parameter for the corresponding week.

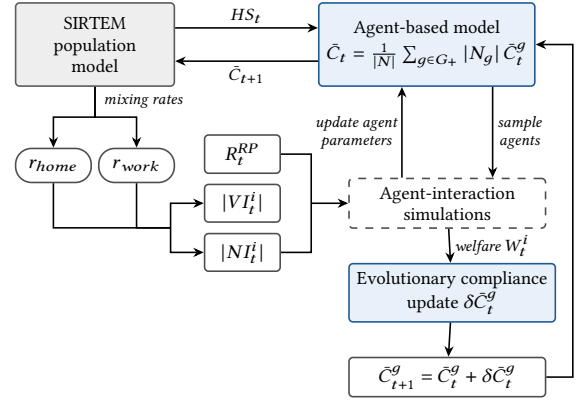
\begin{figure}[t]
    \centering
\begin{tikzpicture}[x=1cm,y=1cm,
  elab/.style={font=\scriptsize\itshape, inner sep=1.5pt, align=center},
  var/.style={wbox, minimum height=0.45cm, minimum width=1.0cm, inner xsep=3pt}]
  \node[gbox, minimum height=1.1cm, minimum width=2.1cm] (sir) at (0.6,0) {SIRTEM\\population\\model};
  \node[box, minimum height=1.1cm, anchor=west] (abm) at (3.55,0)
    {Agent-based model\\[1pt] $\bar{C}_t=\frac{1}{|N|}\sum_{g\in G_+}|N_g|\,\bar{C}^g_t$};
  \draw[arr] ($(sir.east)+(0,0.22)$) -- node[elab, above] {$HS_t$} ($(abm.west)+(0,0.22)$);
  \draw[arr] ($(abm.west)+(0,-0.22)$) -- node[elab, below] {$\bar{C}_{t+1}$} ($(sir.east)+(0,-0.22)$);

  \node[var, rounded corners=6pt] (rh) at (0.05,-1.6) {$\athomerate$};
  \node[var, rounded corners=6pt] (rw) at (1.15,-1.6) {$\atworkrate$};
  \draw[arr] (sir.south) -- ++(0,-0.4) -| (rh.north);
  \draw[arr] (sir.south) -- ++(0,-0.4) -| (rw.north);
  \node[elab, anchor=west] at ($(sir.south)+(0.05,-0.18)$) {mixing rates};

  \node[var] (R)  at (2.55,-1.45) {$R^{RP}_t$};
  \node[var] (VI) at (2.55,-2.1)  {$|VI^i_t|$};
  \node[var] (NI) at (2.55,-2.75) {$|NI^i_t|$};
  \coordinate (bus) at (1.6,-2.425);
  \draw (rh.south) |- (bus);
  \draw (rw.south) |- (bus);
  \draw[arr] (bus) |- (VI.west);
  \draw[arr] (bus) |- (NI.west);

  \node[dbox, minimum height=0.55cm, anchor=west, minimum width=2.6cm] (sim) at (3.95,-2.1) {Agent-interaction\\simulations};
  \node[box, anchor=north west, minimum width=2.6cm] (evo) at ($(sim.south west)+(0,-0.45)$) {Evolutionary compliance\\update $\delta\bar{C}^g_t$};
  \node[wbox, anchor=north west, minimum width=2.6cm] (upd) at ($(evo.south west)+(0,-0.45)$) {$\bar{C}^g_{t+1}=\bar{C}^g_t+\delta\bar{C}^g_t$};

  \draw (R.east) -- ++(0.4,0) |- (NI.east);
  \draw[arr] ($(R.east)+(0.4,-0.65)$) -- (sim.west);
  \coordinate (sa) at ($(sim.north)+(0.6,0)$);
  \coordinate (ua) at ($(sim.north)+(-0.6,0)$);
  \draw[arr] (sa |- abm.south) -- node[elab, right] {sample\\agents} (sa);
  \draw[arr] (ua) -- node[elab, left] {update agent\\parameters} (ua |- abm.south);
  \draw[arr] (sim) -- node[elab, right] {welfare $W^i_t$} (evo);
  \draw[arr] (evo) -- (upd);
  \draw[arr] (upd.east) -- ++(0.35,0) |- (abm.east);
\end{tikzpicture}
    \caption{Coupling between SIRTEM and the ABM. SIRTEM supplies the health-state distribution $HS_t$ and the mixing rates; the ABM samples agents, simulates their interactions, and returns the mean compliance $\totalmeancompliancenext$.}
    \Description{Interaction between the SIRTEM and ABM models.}
    \label{fig:abm-sirtem-model}
\end{figure}

Equations~\eqref{eq:diff_sirtem} and~\eqref{eq:diff_abm} describe the integrated model. It operates daily, alternating between SIRTEM and the ABM and updating their inputs as shown in Figure~\ref{fig:abm-sirtem-model}.
\begin{enumerate}[leftmargin=*,labelsep=0.5em]
    \item \textbf{SIRTEM initialization:} Initialize the health-state distribution $HS_0$ with 20 infected individuals and the remainder susceptible. Set all other health-state counts to zero.
    \item \textbf{ABM initialization:} Initialize the occupation--health group compliance means as $\bar{C}^g_0=0.5$ for all $g\in G$.
    \item \textbf{ABM-SIRTEM iteration:} At time $t$, run SIRTEM for one day to compute $HS_{t+1}$. Then run the ABM iteration in Algorithm~\ref{alg:abm}, using this updated health-state distribution to obtain $\totalmeancompliancenext$.
    \item \textbf{State and compliance updates:} Use the updated occupation--health group compliance means $\meancomplianceattnext$ in the ABM iteration at time $t+1$. Use $\totalmeancompliancenext$ to compute $\beta_{t+1}$ according to Equation~\eqref{eq:sirtem_infection_update}, and pass this infection rate to SIRTEM.
\end{enumerate}
This sequence couples disease dynamics and agent behavior: health states affect the payoffs from interactions, payoffs affect compliance, and compliance feeds back into SIRTEM's infection rate.

\subsection{Bayesian Optimization of Transmission Rates}
\label{sec:bayes}
We estimate $\tau^{tr}$ using Bayesian optimization~\cite{frazier2018tutorial} to minimize the normalized mean squared error (MSE) between simulated and observed daily positive and negative test counts over a $T$-day horizon:
\begin{equation}
\begin{split}
J(\tau^{tr}) = \frac{1}{2T}\sum_{t=1}^{T}\Bigg[
&\left(\frac{\hat{y}_t^{(+)}(\tau^{tr})-y_t^{(+)}}{\bar{y}^{(+)}}\right)^2 \\
+{}&\left(\frac{\hat{y}_t^{(-)}(\tau^{tr})-y_t^{(-)}}{\bar{y}^{(-)}}\right)^2
\Bigg].
\end{split}
\label{eq:mse}
\end{equation}
Here, $\hat{y}_t^{(+)}(\tau^{tr})$ and $\hat{y}_t^{(-)}(\tau^{tr})$ are the model's simulated positive and negative test counts on day $t$. The corresponding observations, $y_t^{(+)}$ and $y_t^{(-)}$, are obtained from the COVID Tracking Project.\footnote{\url{https://covidtracking.com/data}} The normalizing constants are the average observed counts over the evaluation horizon:
\[
\bar{y}^{(\pm)}=\frac{1}{T}\sum_{t=1}^{T}y_t^{(\pm)}.
\]

We optimize over a receding horizon of $K$ weeks. At week $w$, the candidate horizon contains $(\tau^{tr}_{w+k})_{k=0}^{K-1}$. Given candidate initial rates $\tau^{tr}_w$ and $\tau^{tr}_{w+1}$, subsequent rates follow an AR(2) recurrence with coefficients $a,b\in\mathbb{R}$:
\begin{equation}
\tau^{tr}_{w+k}=a\tau^{tr}_{w+k-1}+b\tau^{tr}_{w+k-2},
\qquad k=2,\ldots,K-1.
\label{eq:transmission_ar2}
\end{equation}
We simulate each candidate horizon and evaluate Equation~\eqref{eq:mse} over the corresponding days. These evaluations are used to fit a surrogate model of the objective. New candidates maximize the surrogate model's expected improvement. The candidate with the lowest evaluated objective provides the transmission parameter for week $w$, after which the procedure advances by one week. Algorithm~\ref{alg:bayes_opt} summarizes the process.

\begin{algorithm}
\caption{Bayesian Optimization to Estimate $\tau^{tr}$}
\label{alg:bayes_opt}
\KwInput{Number of weeks $W$; initial samples $n_{\mathrm{init}}$; lookahead horizon $K\geq2$; additional evaluation budget $B$}
\For{$w=0,\ldots,W-1$}{
    Randomly sample $n_{\mathrm{init}}$ candidate tuples $x=(\tau^{tr}_w,\tau^{tr}_{w+1},a,b)$\;
    For each $x$, construct the $K$-week horizon $H(x)$ using Equation~\eqref{eq:transmission_ar2}\;
    Evaluate $J(H(x))$ by simulating ABM-SIRTEM over the evaluation horizon\;
    Fit a surrogate model $M$ to the candidate--objective pairs $(x,J(H(x)))$\;
    \For{$n=1,\ldots,B$}{
        Select $x\in\argmax_x \mathrm{EI}(x;M)$\;
        \tcc{EI denotes expected improvement.}
        Construct $H(x)$ and evaluate $J(H(x))$\;
        Update $M$ with $(x,J(H(x)))$\;
    }
    Retain $\tau^{tr}_w$ from the candidate with the lowest evaluated objective\;
}
\KwOutput{Estimated transmission parameters $(\tau^{tr}_w)_{w=0}^{W-1}$}
\end{algorithm}
\section{Experiments}
\begin{table}[t]
    \centering
    \small
    \setlength{\tabcolsep}{3pt}
    \begin{tabular}{@{}cccrr@{}}
        \toprule
        \textbf{Wave} & \textbf{Start} & \textbf{End} & \textbf{Home} & \textbf{Work} \\
        \midrule
        0 & 22 Mar 2020 & 9 Apr 2020 & 1.5 & 1.38 \\
        1 & 10 Apr 2020 & 16 Jun 2020 & 3.58 & 1.73 \\
        2 & 17 Jun 2020 & 10 Sep 2020 & 3 & 1.81 \\
        3 & 11 Sep 2020 & 1 Mar 2021 & 3.81 & 1.86 \\
        \bottomrule
    \end{tabular}
    \caption{Average daily contacts per person used for Arizona, derived from~\cite{feehan2021quantifying}. Dates indicate the periods over which each wave's estimates are applied in the model.}
    \label{tab:arizona_work}
\end{table}

\begin{figure*}[t]
    \centering
    \includegraphics[width=\textwidth]{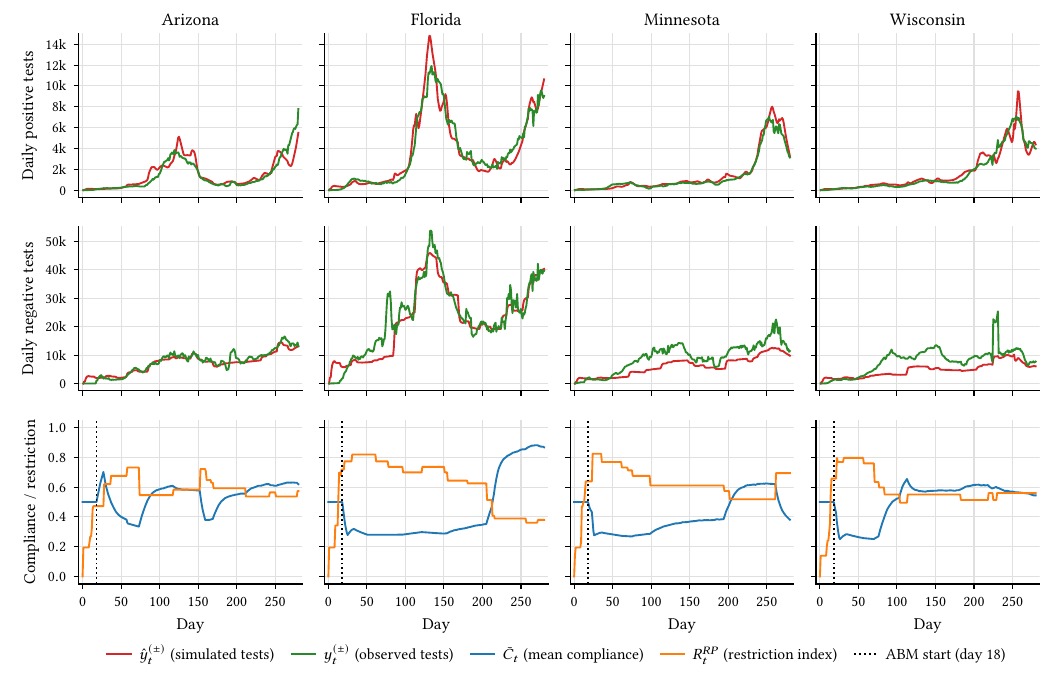}
    \caption{Calibration results for Arizona, Florida, Minnesota, and Wisconsin (columns). Top and middle rows: simulated ($\hat{y}^{(\pm)}_t$) and observed ($y^{(\pm)}_t$) daily positive and negative test counts. Bottom row: simulated mean population compliance $\totalmeancompliance$ and the observed government restriction index $\govtrestriction$; the dotted line marks the day on which the ABM starts. The results for each state were produced by the same calibration procedure (Section~\ref{sec:bayes}).}
    \Description{Twelve panels arranged in three rows and four columns. Each column is a state. The top and middle rows compare simulated and observed positive and negative test counts; the bottom row shows simulated compliance together with the observed restriction index.}
    \label{fig:comparison_all}
\end{figure*}

We calibrate the transmission parameters $\tau^{tr}$ for four U.S. states: Arizona, Florida, Minnesota, and Wisconsin. For each state, we estimate the occupation distribution, set the restriction index $\govtrestriction$, and calculate the at-home and at-work mixing rates, $\athomerate$ and $\atworkrate$, using publicly available data. We estimate $\tau^{tr}$ by Bayesian optimization as described in Section~\ref{sec:bayes}. Figure~\ref{fig:comparison_all} shows the fitted test counts, simulated daily compliance, and observed restriction index. The implementation details are as follows.\footnote{The code repository is planned for release upon publication.}

\begin{enumerate}[leftmargin=*,labelsep=0.5em]
    \item \textbf{Occupation distribution:} We use May 2019 employment estimates from the U.S. Bureau of Labor Statistics (BLS)~\cite{bls_oes_az_2019} to obtain the number of workers in each state by Standard Occupational Classification group. We use telework responses from the October 2022 BLS Current Population Survey~\cite{bls_telework} to estimate the fraction of each occupational group capable of remote work. Specifically, we use the fraction reporting at least some hours of telework as a proxy for remote-work capability. We then use Arizona Executive Order 2020-12~\cite{az_governor}, which defines essential services, to classify the occupational groups as essential or non-essential. Table~\ref{tab:occupation_distribution} gives the estimated occupation distribution for Arizona. Essential services that can be performed remotely are included in the remote category.
    \item \textbf{Restriction index:} We use the Oxford COVID-19 Government Response Tracker (OxCGRT) stringency index~\cite{oxfordcoronavirus,oxford_covid_tracker}, rescaled to $[0,1]$, as the government restriction index $\govtrestriction$. This index summarizes containment and closure policies, including school and workplace closures, restrictions on public gatherings and internal movement, and international travel controls.
    \item \textbf{At-home and at-work mixing rates:} We derive daily average contacts at home and at work from the contact surveys in~\cite{feehan2021quantifying}, which record the relationship between interacting individuals and the locations of their interactions. Table~\ref{tab:arizona_work} lists the averages used for Arizona. We divide the daily averages by the population of the relevant state (e.g., 7.278 million for Arizona) to obtain $\athomerate$ and $\atworkrate$.
    \item \textbf{Agent sampling for interactions:} After calculating the required numbers of necessary and voluntary interactions, $|\necessaryinteractionsi|$ and $|\voluntaryinteractionsi|$, we sample agent pairs at random until the interaction requirements are met, with no repetition within either interaction set. A pair may appear once in each of the necessary and voluntary interaction sets.
    \item \textbf{Time intervals:} Both SIRTEM and the ABM use daily updates. SIRTEM starts on March 4, 2020, and the ABM starts on March 22, 2020, when the contact data needed to calculate $\athomerate$ and $\atworkrate$ become available.
    \item \textbf{Transmission parameters:} We use Optuna~\cite{optuna_2019} for Bayesian optimization of the transmission parameters.
\end{enumerate}

\begin{table}[h]
    \centering
    \small
    \setlength{\tabcolsep}{3pt}
    \renewcommand{\arraystretch}{1.1}
    \begin{tabular}{@{}p{0.25\linewidth}p{0.18\linewidth}p{0.48\linewidth}@{}}
        \toprule
        \textbf{Occupation} & \textbf{Share} & \textbf{Examples} \\
        \midrule
        Essential & 56.26\% & Healthcare, food preparation, legal services \\
        Non-essential, non-remote & 28.13\% & Entertainment, sales \\
        Remote & 15.61\% & IT, business, management \\
        \bottomrule
    \end{tabular}
    \caption{Estimated occupation distribution used for Arizona.}
    \label{tab:occupation_distribution}
\end{table}
\section{Results}
We compare the simulated daily positive and negative test counts obtained with the calibrated transmission parameters against observations for Arizona, Florida, Minnesota, and Wisconsin. The top and middle rows of Figure~\ref{fig:comparison_all} show daily positive and negative counts, respectively. The positive-test trajectories broadly reproduce the timing of the larger observed waves. The fit to negative-test counts varies across states, with substantial underestimation during parts of the Minnesota and Wisconsin simulations. Because observed data within the lookahead horizon are used to estimate the transmission parameters, these comparisons assess calibration fit rather than performance on held-out data.

\textbf{Compliance and restrictions.} The bottom row of Figure~\ref{fig:comparison_all} compares simulated mean compliance with the observed government restriction index. The trajectories show changes in compliance alongside changes in restrictions and test counts. The restriction index is supplied from historical data, so these comparisons do not establish that simulated infections cause policy changes or that changes in compliance cause the observed case trends.
\section{Discussion}
\noindent\textbf{Related Work.} Related research addresses pandemic modeling,
economic impacts, and policy optimization. Bhardwaj et al.
\cite{bhardwaj2020robust} use reinforcement learning to design lockdown
strategies that balance public health objectives and economic costs.
OpenABM-Covid19 \cite{hinch2021openabm} models non-pharmaceutical interventions,
including contact tracing, using individual agents and contact networks.
Duarte et al. \cite{Duarte2021} review studies that integrate economics and
epidemiology to inform pandemic policy, while Dobson et al.
\cite{dobson2023balancing} study the balance between economic and
epidemiological interventions. Hunter et al. \cite{hunter2020hybrid} propose a
hybrid model whose disease component switches between agent-based and
equation-based representations according to the number of infections.
ABM-SIRTEM contributes a coupling between population-level health dynamics and
occupation-dependent welfare incentives that drive changes in agent compliance.

\noindent\textbf{Limitations and Future Work.} The usefulness of the model
depends on the assumptions and parameters used to represent a real population.
We derive several inputs from observed data, but key hyperparameters, including
those in Table~\ref{tab:abm_hyperparameters}, lack readily available ground-truth
values. The welfare components are modeling proxies and have not been validated
against independent economic or well-being measurements. Weekly transmission
rates $\tau^{tr}$ are fitted to historical observations and may absorb effects
that are not explicitly represented in the model. Consequently, agreement with
the calibration data does not establish out-of-sample forecasting performance
or validate the inferred compliance trajectories. Forecasting future outbreaks
requires additional evaluation without access to future observations.

Future work will investigate forecasting under varying conditions and the
optimization of the government restriction index to balance economic welfare
against deaths, hospitalizations, and infections. Formal temporal constraints
could help represent how public health and economic objectives change over
time and support the evaluation of candidate policies.

\noindent\textbf{Conclusions.} We present ABM-SIRTEM to study the socioeconomic
and epidemiological impacts of pandemics. By combining agent-based and
population-level models, we represent individual behavior, occupation-dependent
economic incentives, and feedback between compliance and disease dynamics. We
use multiple data sources to inform model inputs and calibrate transmission
rates against historical positive and negative test counts in four U.S. states.
The resulting simulations illustrate the model's ability to represent coupled
behavioral and epidemiological dynamics, while further validation is needed to
establish its value for forecasting and policy optimization.

\bibliographystyle{ACM-Reference-Format}
\bibliography{sample}
\end{document}